\documentclass[twocolumn,nofootinbib,amsmath]{revtex4}
\usepackage{graphicx}

\begin{document}

\title{Liberating the Inflaton from Primordial Spectrum Constraints}

\author{C. Armend\'ariz-Pic\'on}
\affiliation{Enrico  Fermi Institute and  Department of  Astronomy and
  Astrophysics, \\University of Chicago.}

\begin{abstract}
  I discuss a mechanism that renders the spectral index of the
  primordial spectrum and the inflationary stage independent of each
  other.  If a scalar field acquires an appropriate time-dependent
  mass, it is possible to generate an adiabatic, Gaussian scale
  invariant spectrum of density perturbations during any stage of
  inflation.  As an illustration, I present a simple model where the
  time-dependent mass arises from the coupling of the inflaton to a
  second scalar.  The mechanism I propose might help to implement a
  successful inflationary scenario in particle physics theories that
  do not yield slow-roll potentials.
\end{abstract}

\maketitle

\section{Introduction}
Observations impose  significant constraints on  eventually successful
inflationary models.  Current experimental results are consistent with
a nearly scale invariant spectrum of Gaussian, adiabatic perturbations
\cite{WMAP}.  There is a wide  class of inflationary models that yield
such a  spectrum \cite{inflation}.   In essentially all  these models,
the spectrum is nearly  scale invariant because the universe expansion
closely resembles a de  Sitter stage \cite{MukhanovChibisov}.  In many
cases  however, particularly  when  trying to  embed inflation  within
particle physics theories, it turns out that it is difficult to obtain
quasi  de Sitter inflation,  either because  potentials are  too steep
\cite{SteinhardtBrustein}  or because  the slow-roll  regime  does not
overlap with the regime where the theory is under control \cite{Lyth}.
 
Unlike the  slope of the  primordial spectrum, its amplitude  does not
necessarily depend  on the  inflationary epoch itself.   If primordial
perturbations  originate  from  the  decay  of  a  ``curvaton''  field
\cite{curvaton},  or from the  fluctuating couplings  ``constants'' of
the inflaton  \cite{DGZ, Kofman}, the final amplitude  of the spectrum
turns  out not  to be  directly related  to  inflation.  Nevertheless,
these scenarios still  had to contain a stage  of de Sitter inflation,
in  order  for the  perturbations  in  the  curvaton or  the  coupling
constants of the inflaton to be to scale invariant.

In this paper,  I propose a mechanism that  additionally decouples the
spectral index  from the physics  of the inflaton.  In  this scenario,
fluctuations  are imprinted  on  a ``test''  scalar  field whose  mass
changes with  time.  The time-varying  mass reproduces the  effects of
gravity  during a de  Sitter stage,  even though  the universe  is not
expanding   exponentially  fast.    At  the   end  of   inflation  the
fluctuations imprinted in the test  field are transferred to the decay
products of the inflaton by  the mechanism proposed by Dvali, Gruzinov
and  Zaldarriaga \cite{DGZ}.   In that  way,  the liberated inflaton
does not have to satisfy constraints from the amplitude
and slope of the primordial spectrum.

The  paper is organized  I follows.   In Section  \ref{sec:varying}, I
describe how  a scalar with  a time-varying mass  can lead to  a scale
invariant  spectrum of primordial  density perturbations.   In Section
III,  I present  an example  where  the changing  mass is  due to  the
coupling   to  the   scalar   that  drives   inflation.   In   Section
\ref{sec:non-inflating},  I   try  to   extend  the  mechanism   to  a
non-inflating  universe, and in  Section \ref{sec:conclusions}  I draw
the conclusions.

\section{Time-varying mass}\label{sec:varying}

Consider a test scalar field  $\varphi$ with mass $m$ in an expanding,
flat, linearly perturbed Friedmann-Robertson-Walker universe,
\begin{equation}
  ds^2=a^2(\eta)\left[(1+2\Phi)d\eta^2-(1-2\Phi)d\vec{x}^2\right].
\end{equation}
Here, $a$ is  the scale factor and $\Phi$  the gravitational potential
(in   longitudinal   gauge).   Neglecting   for  the   moment   metric
perturbations, the scalar field equation of motion is
\begin{equation}\label{eq:motion}
  v_k''+\left(k^2+m^2 a^2-\frac{a''}{a}\right)v_k=0,
\end{equation}
where $v=a\,  \varphi$, a prime  denotes a derivative with  respect to
conformal time  $\eta$, and the  subindex $k$ denotes the  $k$ Fourier
component.   Because  Eq.  (\ref{eq:motion})  is  linear  in $v$,  the
equation of motion for $\varphi$ and its perturbations $\delta\varphi$
agree.

For  simplicity, let  me  momentarily consider  a power-law  inflating
universe,
\begin{equation}\label{eq:a}
  a\propto|\eta|^\frac{\beta}{1-\beta}.
\end{equation}
In cosmic time, the last equation corresponds to ${a\propto t^\beta}$.
Hence, the universe inflates for $\beta>1$, and in that case conformal
time $\eta$ runs  from $-\infty$ to $0$. Suppose  now that the squared
mass of the field is proportional to the squared Hubble parameter,
\begin{equation}\label{eq:mass}
  m^2= c\cdot H^2,
\end{equation}
where $c$ is a  (dimensionless) constant coefficient.  Such a relation
simply arises for  instance if the scalar is  non-minimally coupled to
gravity,
\begin{equation}
  L_\varphi=\frac{1}{2}\partial_\mu \varphi \partial^\mu \varphi
    +\frac{c\cdot \beta}{12(2\beta-1)}\, R \, \varphi^2,
\end{equation}
or, as I  discuss in Section \ref{sec:example}, it  can arise from the
coupling of $\varphi$ to a second scalar field $\chi$,
\begin{equation}
  L_\varphi=\frac{1}{2}\partial_\mu \varphi \partial^\mu \varphi
    -\frac{1}{2}m^2(\chi)\varphi^2.
\end{equation}
Also, it  has been observed  that supersymmetry breaking in  the early
universe  induces scalar  field  masses  of the  order  of the  Hubble
parameter  along  flat  directions  \cite{DiRaTh}.   For  our  present
purposes though,  it will suffice  to treat Eq.  (\ref{eq:mass})  as a
phenomenological relation.

If  $a$ is  given by  Eq.  (\ref{eq:a}),  and $m^2$  is given  by  Eq. 
(\ref{eq:mass}) the equation of motion (\ref{eq:motion}) reads
\begin{equation}\label{eq:motionBessel}
  v_k''+\left(k^2-\frac{\nu^2-1/4}{\eta^2}\right)v_k=0,
\end{equation}
where
\begin{equation}\label{eq:nu}
  \nu=\frac{\sqrt{(9-4c)\beta^2-6\beta+1}}{2(\beta-1)}.
\end{equation}
The   solution  of   Eq.   (\ref{eq:motionBessel})   with  appropriate
``adiabatic vacuum'' initial conditions \cite{vacuum} is
\begin{equation}\label{eq:solution}
  v_k=\sqrt{\frac{\pi (-\eta)}{2}} H_\nu(-k \eta),
\end{equation}
where $H_\nu$  is the Hankel function  of the first  kind. Because the
fluctuations   in   $\varphi$    arise   from   vacuum   fluctuations,
$\delta\varphi_k$ is a Gaussian variable.

The  power spectrum  $\mathcal{P}_\varphi$ is  a measure  of  the mean
square fluctuations  of $\varphi$ on comoving  lengthscales $1/k$, and
it is defined by \cite{MuFeBr}
\begin{equation}
  \mathcal{P}_\varphi=\frac{k^3}{4\pi^2} \frac{|v_k|^2}{a^2}.
\end{equation}
Cosmologically relevant modes are larger than the Hubble radius at the
end of inflation. In this long-wavelength limit, the power spectrum is
then
\begin{equation}\label{eq:power}
  \mathcal{P}_\varphi=\frac{2^{2\nu} |\Gamma(\nu)|^2}{8\pi^3}
  \left(\frac{\beta-1}{\beta}\right)^{2\nu-1}\, 
  H^2 \cdot\left[\left(\frac{H}{H_*}\right)^{\beta-1}
    \frac{k}{k_*}\right]^{n_s-1}.
\end{equation}
Note  that  the amplitude  of  the  spectrum  is time-dependent.   The
comoving  scale  $k_*$ is  an  arbitrary  reference  scale, and  $H_*$
denotes the value  of the Hubble constant when  that scale crosses the
Hubble radius, $a/k_*=H_*^{-1}$.  In  the following, I denote by $k_*$
the scale that corresponds to  our present Hubble radius. The spectral
index is
\begin{equation}\label{eq:index}
n_s-1=3-2\nu.
\end{equation}
A scale invariant Harrison-Zeldovich spectrum, corresponds to $n_s=1$.
Hence, from Eqs. (\ref{eq:nu}) and (\ref{eq:index}) scale invariance
requires
\begin{equation}\label{eq:c}
  c=\frac{3\beta-2}{\beta^2}.
\end{equation}
Because cosmic microwave background anisotropies \cite{WMAP} limit the
departures from  $n_s=1$ to  less than about  ten per cent,  for given
$\beta$ or order one, $c$  has to agree with Eq.  (\ref{eq:c}) roughly
to that  accuracy.  Therefore,  from Eq. (\ref{eq:mass}),  the squared
mass  is  positive during  power-law  inflation, negative  (tachyonic)
during  pole-like inflation  and vanishing  for de  Sitter  inflation. 
Here, I restrict myself to $1<  \beta\leq \infty$.  If $c$ is given by
Eq.  (\ref{eq:c}) the power spectrum is then
\begin{equation}\label{eq:invpower}
  \mathcal{P}_\varphi=\left(\frac{\beta-1}{\beta}\right)^2
  \frac{H^2}{4\pi^2}.
\end{equation}
Thus, whereas  the fluctuations of  $\varphi$ in de Sitter  approach a
constant value  $H/2\pi$, they  decay as $a^{-1/\beta}$  if $\beta\neq
\infty$.  

The  mechanism I have  just described  successfully generates  a scale
invariant spectrum of perturbations  in $\varphi$, which is \emph{not}
what  is  required.   Current  experiments  favor  a  scale  invariant
spectrum of Gaussian, adiabatic, \emph{density perturbations}.  Hence,
perturbations in $\varphi$ need  to be transferred to perturbations in
the radiation produced at the end of inflaton.  As I mention in detail
in  next Section,  this can  be  accomplished if  the field  $\varphi$
determines the value of the couplings constants of the inflaton to its
decay products  \cite{DGZ, Kofman}.  Of course,  if primordial density
perturbations  originate   form  fluctuations  in   $\varphi$,  it  is
important  that perturbations  due  to the  inflaton be  significantly
smaller than the ones due  to the spatial variation of $\varphi$.  The
power  spectra of scalar  metric perturbations  $\mathcal{P}_\Phi$ and
gravitational waves  $\mathcal{P}_h$ seeded during  inflation are (see
for instance \cite{vacuum})
\begin{equation}
  \mathcal{P}^{inf}_{\Phi}\sim \beta\, \frac{H_*^2}{M_{Pl}^2}
  \left(\frac{k}{k_*}\right)^{-\frac{2}{\beta-1}}
  \sim \beta\, \mathcal{P}^{inf}_h.
\end{equation}
Because both spectra  are red, the highest amplitude  in an observable
mode  is attained  for the  present horizon  $k_*$.   Cosmic microwave
measurements  have determined  that  $\mathcal{P}_\Phi\sim 10^{-10}$.  
Hence, in  order for scalar perturbations seeded  during inflation not
to  account for  the observed  anisotropies, cosmic  inflation  has to
occur at a low energy scale,
\begin{equation}\label{eq:constraint}
  \beta\frac{H_*^2}{M_{Pl}^2}\ll 10^{-10}.
\end{equation}
This also implies that gravitational waves have a negligible impact on
the cosmic microwave background.

Although for  simplicity I have  focused on power-law  inflation, this
scenario can be easily generalized to any epoch of inflation.  Suppose
that an  arbitrary $a(t)$ is given. All  we need is that  in the given
FRW spacetime there is no particle horizon. Then, the integral
\begin{equation}\label{eq:integral}
  \eta_e-\eta\equiv \int_{t}^{t_e}\frac{d\tilde{t}}{a(\tilde{t})}
\end{equation}
diverges  as  $t$ approaches  cosmic  time  origin $t_i$.   Therefore,
conformal time runs  from $-\infty$ to $\eta_e$, where  $\eta_e$ is an
arbitrary end time  which I shall identify with the  end of inflation. 
In  such a  spacetime there  is an  apparent ``event  horizon'',  i.e. 
light emitted at time $t$ can reach an observer before time $t_e$ only
if it is emitted within a physical distance
\begin{equation}
  d_E=a(\eta)(\eta_e-\eta).
\end{equation}
Note that this  apparent event horizon is not in  general a real event
horizon, since the upper  limit in the integral (\ref{eq:integral}) is
kept fixed and  finite.  The apparent event horizon  is not the Hubble
radius  $H^{-1}$ either,  though in  many cases  they are  roughly the
same.

Suppose now that the mass of a scalar field is given by
\begin{equation}\label{eq:mass1}
  m^2=\frac{a''}{a^3}-\frac{2}{d_E^2}.
\end{equation}
Because there is an arbitrary  freedom in the coupling of $\varphi$ to
a second  scalar field, such an  evolution of the squared  mass can be
always  achieved.    Substituting  Eq.   (\ref{eq:mass1})   into  Eq.  
(\ref{eq:motion}) I get
\begin{equation}\label{eq:motion1}
  v_k''+\left(k^2-\frac{2}{(\eta_e-\eta)^2}\right)v_k=0.
\end{equation}
A conformal  time shift in  Eq.  (\ref{eq:motionBessel}) leads to  Eq. 
(\ref{eq:motion1}). Hence,  the solutions  of the latter  equation are
given   by  Eq.    (\ref{eq:solution}),  with   $-\eta$   replaced  by
$\eta_e-\eta$.   At early  times,  $\eta_e-\eta$ approaches  infinity,
i.e.  all modes  are initially within the horizon,  $a/k \ll d_E$.  At
late  times,  $\eta_e-\eta$ approaches  zero  and  all  the modes  are
super-horizon sized, $a/k\gg d_E$.  In this long-wavelength limit, the
spectral index  is still given  by Eq.  (\ref{eq:index}),  where, from
Eqs.   (\ref{eq:motion1})   and  (\ref{eq:motionBessel}),  $\nu=3/2$.  
Again, this corresponds to a scale invariant spectrum.

\section{A concrete example}\label{sec:example}

It  might  seem  that  the  procedure  described  above  is  extremely
fine-tuned  since,  according   to  Eq.   (\ref{eq:mass1}),  the  time
evolution of  a squared  mass has to  accurately reflect  the a-priori
independent  expansion  history.  However,  in  the  presence of  (non
slow-roll)  inflationary attractors,  it turns  out that  the required
values of the squared mass come about surprisingly naturally.

Consider for  instance two  coupled scalar fields  in the  presence of
Einstein gravity,
\begin{eqnarray}\label{eq:action}
\int d^4x \sqrt{-g}\Bigg[-\frac{M_{Pl}^2}{16\pi}R
+\frac{1}{2}\partial_\mu \varphi 
    \partial^\mu\varphi+\frac{1}{2}\partial_\mu \chi 
    \partial^\mu \chi- \\ \nonumber
    {}-\left(1+\frac{4\pi\, \beta\,c}{3\beta-1}\cdot \
      \frac{\varphi^2}{M_{Pl}^2}\right)
    V(\chi)+L_m\Bigg].
\end{eqnarray}
Here,  $M_{Pl}=G^{-1/2}\approx 10^{19}\,  \mathrm{GeV}$ is  the Planck
mass, and
\begin{equation}\label{eq:vofchi}
  V(\chi)=V_0\exp\left(-\sqrt{\frac{16\pi}{\beta}}
   \frac{\chi}{M_{Pl}}\right).
\end{equation}
In  our  example,  the  field   $\varphi$  is  assumed  to  remain  at
$\varphi=0$.   Hence,  it  is   sufficient  to  consider  a  quadratic
$\varphi^2$ term  in the  action (\ref{eq:action}).  The  inclusion of
higher  even powers of  $\varphi$ in  the action  will not  change our
results\footnote{As long  as the coefficients  of those terms  are not
  too large.}.  Exponential  potentials and couplings naturally appear
in string  theory (from  the dilaton), and/or  in theories  with extra
dimensions (from  the radion).  $L_m$  stands for additional
matter terms only  relevant during reheating; I shall  write them down
below.

Suppose  that  initially  the  field  $\varphi$  sits  at  the  origin
$\varphi=0$.   Because the  squared  mass of  the  field is  positive,
$\varphi=0$  is a  stable solution  of the  background  equations.  If
$\varphi=0$, the field $\chi$  effectively evolves in the single field
potential  (\ref{eq:vofchi}).   These  potentials  are known  to  have
power-law  inflationary attractors  \cite{power-law}, along  which the
expansion of the universe is given by Eq. (\ref{eq:a}) and
\begin{equation}\label{eq:attractor}
  \frac{V(\chi)}{M_{Pl}^2}=\frac{3\beta-1}{8\pi \beta}H^2.
\end{equation}
The   coefficient   in   front    of   $\varphi^2$   in   the   action
(\ref{eq:action})   implies   that   the   field   $\varphi$   has   a
$\chi$-dependent mass
\begin{equation}\label{eq:phimass}
  m^2_\varphi(\chi)=\frac{8\pi\,\beta\,c}{3\beta-1}
  \frac{V(\chi)}{M_{Pl}^2}.
\end{equation}
Thus, from Eqs. (\ref{eq:phimass}) and (\ref{eq:attractor}), along the
inflationary  attractor the  squared  mass of  $\varphi$ is  precisely
given by Eq.  (\ref{eq:mass}).

The    linearized     equations    for    the     perturbations    are
\cite{MukhanovSteinhardt}
\begin{eqnarray}
  \delta\varphi''&+&2\mathcal{H}\delta\varphi'+k^2\delta\varphi
  -4\varphi'\,\Phi' + 2\, m_\varphi^2\, a^2\, \varphi\,\Phi+ \nonumber \\ 
  {}&+&m_\varphi^2\,a^2 \,\delta\varphi
  +\frac{d m_\varphi^2}{d\chi}\, a^2 \, \varphi \, \delta\chi=0,
  \label{eq:perturbed}\\  
  \delta\chi''&+&2\mathcal{H}\delta\chi'+k^2\delta\chi
  -4\chi'\Phi'+ 2\frac{dV}{d\chi}a^2\, \Phi+ \nonumber \\ 
  {}&+&\frac{dm_\varphi^2}{d\chi}\,a^2\,\varphi^2\,\Phi
  +\frac{d^2V}{d\chi^2}\, a^2 \, \delta\chi+ \nonumber \\
  {}&+&\frac{1}{2}\frac{d^2 m_\varphi^2}{d\chi^2}\, a^2 \,\varphi^2\,
  \delta\chi
    +\frac{dm_\varphi^2}{d\chi}\, a^2 \, \varphi\, \delta\,\varphi=0, \\
  \Phi'&+&\mathcal{H}\Phi=\frac{4\pi}{M_{Pl}^2}
  \left(\varphi'\delta\varphi+\chi'\delta\chi'\right),
\end{eqnarray}
where $\mathcal{H}\equiv a'/a$.   Consequently, to linear order around
the solution  $\varphi=\varphi'=0$, perturbations in  $\varphi$ do not
couple  to  inflaton  or  metric  perturbations and  vice  versa.   In
particular,     substituting    $\delta\varphi=v/a$    into     Eq.    
(\ref{eq:perturbed}) yields  Eq.  (\ref{eq:motion}).  In  summary, our
model satisfies the assumptions made in Section \ref{sec:varying}.

If the  inflaton potential is globally given  by Eq. (\ref{eq:vofchi})
inflation  never  ends.   As  in conventional  power-law  inflationary
models,  I  shall assume  that  $V(\chi)$  develops  a minimum  around
$\chi=\chi_{end}$.   Hence,  when   $\chi$  reaches  the  vicinity  of
$\chi_{end}$,  inflation  ends  and  $\chi$ starts  oscillating,  thus
reheating the  universe.  During  the oscillating phase,  the universe
evolves as if dominated by dust, and the time average of $V(\chi)$ is
\begin{equation}
  \frac{\langle V(\chi)\rangle}{M_{Pl}^2}=
  \frac{3}{16\pi}\langle H^2\rangle.
\end{equation}
During that time  the squared mass of the  field is still proportional
to  the squared Hubble  parameter and  the shape  of the  spectrum for
modes larger than the Hubble radius remains unaltered.  I shall assume
that the amplitude of the $\delta\varphi$ perturbations hardly changes
during  a  short,  almost  instantaneous stage  of  reheating,  though
parametric resonance  might under certain  circumstances significantly
boost the initial amplitude \cite{BrandenbergerFinelli}.

Reheating occurs due the coupling of the inflaton $\chi$ to additional
matter fields.  In  order to transfer the scale  invariant spectrum of
perturbations imprinted  in $\varphi$ into  the decay products  of the
inflaton,   I   assume,   following   \cite{DGZ},  that   the   action
(\ref{eq:action}) contains terms of the form
\begin{equation}
  L_m=-\lambda_0\left(1+f\frac{\varphi}{M_{Pl}}\right)\,
  \chi \bar{\psi}\psi. 
\end{equation}
Here, the fermion $\psi$ generically  stands for the decay products of
the inflaton,  and $\lambda_0$ and $f$ are  two dimensionless coupling
constants.    The  energy  density   at  the   end  of   reheating  is
\cite{KoLiSt}
\begin{equation}\label{eq:reheating}
  \rho_R\sim T_R^4\sim
  \lambda_0^4\left(1+f\frac{\varphi}{M_{Pl}}\right)^4 
  m_\chi^2 M_{Pl}^2
\end{equation}
where $T_R$ is the reheating  temperature and $m_\chi$ is the inflaton
mass  during   the  oscillating  stage.    Therefore,  because  during
inflation $\varphi=0$,  fluctuations in $\varphi$  produce fluctuations
in the energy density \cite{DGZ}
\begin{equation}\label{eq:amplitude}
  \frac{\delta\rho}{\rho}\sim \frac{\delta T}{T}
  \sim f \frac{\delta\varphi}{M_{Pl}},
\end{equation}
which cosmic microwave  background measurements \cite{WMAP} require to
be are  around $10^{-5}$. Note  that the amplitude of  the temperature
perturbations  is  fixed by  $f$,  whereas  the reheating  temperature
itself  is  determined  by  $\lambda_0$  and $m_\chi$.   For  a  given
reheating  temperature, one  can compute  the number  of  $e$-folds of
inflation $N$ after the present horizon left the Hubble radius,
\begin{equation}
  N\sim\frac{\beta}{\beta-1}\log \left(\frac{T_0 \,T_R}{H_0 M_{Pl}}\right),
\end{equation}
where, $H_0\sim 10^{26}\, \mathrm{m}$ and $T_0\sim 3\, \mathrm{K}$ are
respectively the  present values  of the Hubble  constant and  the CMB
temperature.  During this number of $e$-folds, the seeded modes span a
window in $k$ space
\begin{equation}\label{eq:window}
  \frac{k_{end}}{k_{*}}=\exp\left(\frac{\beta-1}{\beta}N\right)=
  \frac{T_0\, T_R}{H_0 M_{Pl}}.
\end{equation} 
Thus, the  amount of seeded  modes does not  agree with the  amount of
inflation.  The  cosmologically accessible window  $k/k_*$ spans three
to four logarithmic decades.

Observations are  consistent with adiabatic  primordial perturbations,
though  significant  amounts  of  non-adiabaticity, of  the  order  of
several  tenths  per  cent,  are still  compatible  with  observations
\cite{isocurvature}.  Because  in our scenario reheating  is driven by
the oscillations of a single component, the $\chi$ field, the spectrum
of density perturbations produced at the end of reheating is adiabatic
\cite{RiottoMatarrese}.  Current observations also restrict the amount
of non-Gaussianity  of the perturbations.  It  is typically quantified
by             means            of             the            relation
$\Phi=\Phi_g+f_{NL}(\Phi_g^2-\langle\Phi_g^2\rangle)$
\cite{KomatsuSpergel}, which  expresses metric perturbations  in terms
of a  Gaussian random field  $\Phi_g$.  The amount  of non-Gaussianity
produced in  our model was carefully  estimated in \cite{Zaldarriaga},
where it was found that $f_{NL}$  is of order one. This is well within
the experimental limits  $-58<f_{NL}<134$ \cite{Komatsu}, but might be
potentially detectable in the future \cite{KomatsuSpergel}.

To conclude this  section, let me show that, at  least at first sight,
this model is phenomenologically viable.  Assume $m_\chi\sim 10^2$ TeV
and  $\lambda_0\sim  10^{-6}$.   From Eq.   (\ref{eq:reheating}),  the
reheating  temperature is  $T_R=10^3\, \mathrm{TeV}$,  well  above the
nucleosynthesis limit.  Substituting into Eq. (\ref{eq:window}) I find
$k_{end}/k_*\sim  10^{16}$, which  is  much bigger  than the  required
$10^4$.   Let  me  set  $\beta=4$.   For  a  vanishing  mass  ($c=0$),
power-law inflation with such an exponent would yield a spectral index
$n_s=1/3$  \cite{vacuum},  very far  form  the experimentally  favored
scale invariant  spectrum $n_s\approx 1$.  With  a time-dependent mass
and $c=5/8$, the spectrum is scale invariant.  The number of $e$-folds
from the time our present horizon  left the Hubble radius till the end
of inflation is  then $N\sim49$, and the total  number of $e$-folds of
inflation  is larger.  This  suffices to  explain the  homogeneity and
flatness   of    the   universe.    Let   me    choose   in   addition
$V(\chi_{end})\sim m_\chi^4$.   Then, the value of  the squared Hubble
parameter at crossing was $H_*^2\sim10^{-45}M_{Pl}^2$, which satisfies
the  constraint  (\ref{eq:constraint}).   If  in  addition  we  choose
$\lambda_0 f/M_{Pl}\sim 10^2/m_\chi$ we find that (\ref{eq:amplitude})
reproduces the observed value of the primordial spectrum amplitude.

\section{Non-Inflating Universe}\label{sec:non-inflating}

In  conventional  inflationary  models, primordial  perturbations  are
causally seeded when modes exit  the sound horizon (which for a scalar
field is of  the order of the Hubble radius).  In  the scenario I have
previously  described,  the  Hubble  radius  and  the  Compton  radius
$m^{-1}$ are  of the  same order,  so one could  argue that  modes are
causally  seeded because  the  Compton radius  grows  slower than  the
physical  wavelength  of   the  perturbations.  Similarly,  one  could
envisage  a scenario  where  perturbations are  causally  seeded in  a
non-inflating  universe because  the Compton  radius starts  large and
subsequently does not grow fast enough. Let me illustrate this idea in
a concrete setting.

Consider this  time an  expanding, non-inflating universe.   Then, the
integral (\ref{eq:integral})  converges as cosmic  time approaches the
origin $t_i$. In that case, there is a particle horizon
\begin{equation}
d_P=a(\eta) (\eta-\eta_i)=a \int_{t_i}^t 
\frac{d\tilde{t}}{a(\tilde{t})}.
\end{equation}
Suppose that  in analogy with Eq.  (\ref{eq:mass1}),  the squared mass
of a scalar is given by
\begin{equation}\label{eq:mass2}
  m^2=\frac{a''}{a^3}+m_{eff}^2,
\end{equation}
where
\begin{equation}\label{eq:mass3}
m_{eff}^2=-\frac{2}{a^2(\eta_e-\eta)^2}
\end{equation}
and $\eta_e$ is  a constant with dimensions of  length.  As in Section
\ref{sec:example},  such  a  relation  can  be  obtained  by  coupling
$\varphi$  to a second  scalar $\chi$.   The rationale  of considering
this  squared  mass  becomes  manifest  when writing  down  the  field
equation  of motion  (\ref{eq:motion}), which  takes the  form of  Eq. 
(\ref{eq:motion1}). Thus, the equations of motion of the perturbations
both  in an  inflating  as well  as  in a  non-inflating universe  are
formally the same.   As in Section \ref{sec:varying}, this  leads to a
scale invariant spectrum.

There  is a  crucial  difference  though between  an  inflating and  a
non-inflating universe.   In an  inflating universe, $\eta$  runs from
$-\infty$  to  $\eta_e$.  Therefore,  in  Eq.  (\ref{eq:motion1})  the
$k^2$ term dominates at early times (short wavelength regime), and the
$1/(\eta_e-\eta)^2$  term  dominates at  late  times (long  wavelength
regime).   This   is  why  modes   exit  the  Hubble  radius.    In  a
non-inflating universe  though, $\eta$ runs from $0$  to $\eta_e$, and
hence, modes start in the  short-wavelength regime only if $\eta_e$ is
big  enough.    Of  course,  the  difference   between  inflating  and
non-inflating spacetimes rests on the existence or absence of particle
and apparent event horizons.

\begin{figure}
  \begin{center}
    \includegraphics{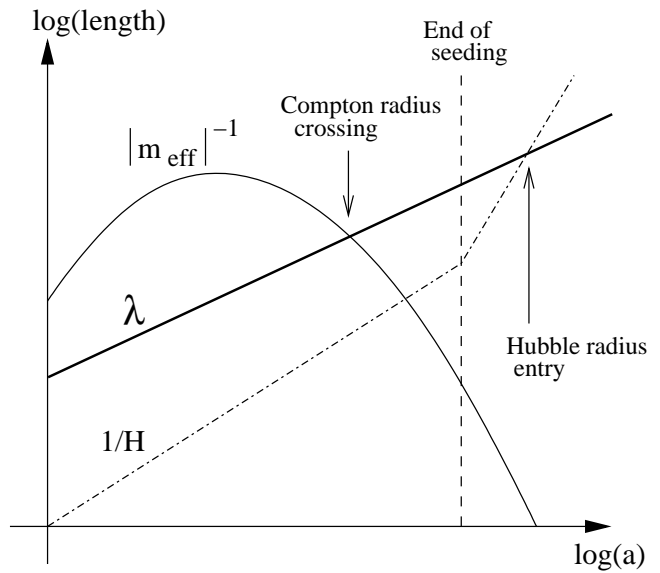}
    \caption{A schematic plot of the evolution of the different
      length scales.   The physical  length of a  given mode  is shown
      with a  thick line.  Initially,  the mode is outside  the Hubble
      radius   (dot-dashed  line)  but   inside  the   Compton  radius
      (continuous line).  As  the mass of the field  evolves, the mode
      exits the  Compton radius, freezes  and later enters  the Hubble
      radius.
      \label{fig:evolution}}
  \end{center}
\end{figure}

The mode  evolution is shown  in Figure \ref{fig:evolution}.   A given
perturbation  mode might  be  initially super-Hubble  sized but  still
within  the Compton radius.   This requires  that the  total effective
mass of  the scalar be small  compared to the  Hubble radius.  Because
the coupling to  gravity generates a (tachyonic) mass  of the order of
the   Hubble    radius,   ${m^2\approx   -   a''/a^3}$    (see   Eq.   
(\ref{eq:motion})),  the  former  requirement  forces  a  cancellation
between that term and the one  coming from the ``true'' mass.  This is
the origin  of the first term  on the r.h.s of  Eq.  (\ref{eq:mass2}). 
Note that  even though this  scenario can seed perturbations  on super
Hubble scales, it does not  violate causality.  The reason is that the
evolution of perturbations on  different lengthscales is dictated by a
time-varying mass which is the  same in the whole universe.  Thus, the
ultimate origin of super Hubble correlations is the homogeneity of the
universe, which, in the absence  of cosmic inflation, we assume rather
than explain.
 
Can  this scale  invariant spectrum  seeded during  a non-inflationary
stage  spectrum account for  cosmologically relevant  modes?  Consider
now  a   stage  of   power-law  expansion,  Eq.    (\ref{eq:a}),  with
$0<\beta<1$.  In that case, Eq. (\ref{eq:mass3}) takes the form
\begin{equation}
  m_{eff}^2=-2\,\frac{(1-\beta)^2}{\beta^2}
  \left[\left(\frac{H}{H_e}\right)^{1-\beta}-1\right]^{-2} H^2,
\end{equation}
where  $H_e$ is  the value  of $H$  when $\eta=\eta_e$  (at  that time
$m_{eff}$ diverges).   Our present comoving horizon is  $k_*=a_0 H_0$. 
Suppose that this mode left the effective Compton radius $N$ $e$-folds
before the end  of seeding, which concluded at  temperature $T_R$. The
viability  of this  scenario requires  that seeding  ends  before time
$\eta_e$.   The  condition   of  effective  Compton  radius  crossing,
$a/k_*=|m_{eff}|^{-1}$, then translates into
\begin{equation}\label{eq:crossing}
  \left(\frac{T_R^2}{H_e M_{Pl}}\right)^{1-\beta}-e^{(1-1/\beta)N}
  \sim\frac{1-\beta}{\beta}
  \frac{T_0^2}{H_0\, M_{Pl}}\frac{T_R}{T_0}.
\end{equation}
Let  us   assume  that  the  seeded  ended   before  nucleosynthesis.  
Nucleosynthesis occurs  around $T \sim 10^{10}\,  T_0$.  Hence, unless
$\beta$ is very close to one, Eq.  (\ref{eq:crossing}) implies
\begin{equation}\label{eq:limit}
  \left(\frac{T_R^2}{H_e M_{Pl}}\right)^{1-\beta}
  \geq 10^{-3} \frac{T_R}{T_0}\geq 10^7\gg 1.
\end{equation}
Using the crossing condition for the mode that left the Compton radius
at the end of inflation, one can derive the amount of seeded modes,
\begin{equation}
  \frac{k_{end}}{k_*}\sim e^N \frac{T_R^2}{H_* M_{Pl}}
  \frac{\left(H_*/H_e \right)^{1-\beta}-1}
  {\left(T_R^2/H_e M_{Pl} \right)^{1-\beta}-1}.
\end{equation}
Evaluating the last expression in the limit (\ref{eq:limit}) I find
\begin{equation}
  \frac{k_{end}}{k_*}\sim 1,
\end{equation}
where  I have  used  the  fact that  energy  density decreases  during
expansion,  $H_*^2\sim  \rho_*/M_{Pl}^2> T_R^4/M_{Pl}^2$.   Therefore,
this scenario  is unable to  explain the origin of  the cosmologically
relevant window  $k/k_*\sim 10^3$.  There  are only two escapes  I can
think of.  The first is to assume that $\beta\approx 1$, but then, the
spacetime  is already at  the verge  of inflating.   The second  is to
assume that  seeding proceeds during a stage  of radiation domination,
$\beta=1/2$.   This   allows  the  end  of  seeding   to  occur  after
nucleosynthesis.  In  that way, the estimate  in Eq.  (\ref{eq:limit})
can avoided simply by setting $T_R/T_0\sim 10^3$, which corresponds to
the temperature around  the time when decoupling occurs.   But at that
time, perturbations  have to be already  in place and  the universe is
rather matter-dominated.   It is doubtful that this  scenario can work
even in such an extreme case.
 
\section{Summary and Conclusions}\label{sec:conclusions}

During   cosmic  expansion,   gravity  contributes   a  time-dependent
correction to the  total effective mass of a  minimally coupled scalar
field, Eq.   (\ref{eq:motion}).  If this total  effective mass evolves
appropriately,  a  scale  invariant  spectrum of  scalar  fluctuations
results,  Eq.   (\ref{eq:invpower}).    The  ultimate  origin  of  the
effective time-dependent  mass is however not very  important.  It can
arise solely  from the expansion  of the universe,  as in a  de Sitter
universe or, as I have explored  in this paper, it can also arise from
the coupling of  the scalar to other evolving  fields. As a particular
example, I  have shown that it  is possible to seed  a scale invariant
spectrum  of perturbations  in  a  scalar field  during  any stage  of
power-law inflation.  A concrete model that realizes this setting, Eq.
(\ref{eq:action}),   relies   on  two   coupled   scalars  and   looks
surprisingly simple. One  of the fields is the  inflaton, which drives
power-law inflation, and  the other is a test  scalar field sitting at
the minimum of  its effective potential, upon which  a scale invariant
spectrum of fluctuations is imprinted.

A  scale invariant  spectrum of  scalar field  perturbations  does not
suffice  however  to account  for  the  observed  spectrum of  density
perturbations. If there was no way to transfer the field perturbations
to energy density perturbations, this scenario would not be realistic.
Recently though, a new mechanism  has been proposed to transfer scalar
field perturbations  to density perturbations  \cite{DGZ, Kofman}.  If
the  couplings  of the  inflaton  to  its  decays products  are  field
dependent,  fluctuations   in  the   latter  can  be   converted  into
fluctuations  in matter  and radiation.   As  any model  based on  the
reheating mechanism of \cite{DGZ}, our scenario predicts a substantial
larger, though  observationally consistent, degree  of non-Gaussianity
in the primordial spectrum.   Also, there is no substantial production
of gravitational waves because  the amplitude of perturbations in the
inflaton are  assumed to be  insufficient to account for  the observed
temperature anisotropies.

The idea I have discussed  can also explain a scale invariant spectrum
of density perturbations during a non-inflationary stage of expansion.
In this  case, modes are seeded  when they cross  an effective Compton
radius  which evolves  in  time.  But  the  generated spectrum  cannot
encompass a sufficient window of modes around the present horizon.  In
this  case,  the  origin  of   large  (but  not  large  enough)  scale
correlations can  be traced back  to the homogeneity of  the universe,
which is assumed, rather than explained.

In summary, in the mechanism  I have described the primordial spectrum
does  not  depend on  the  nature of  the  inflationary  stage.  As  a
consequence, the  inflaton does  not have to  account for  neither the
amplitude nor  the spectral index of the  primordial spectrum.  Hence,
``liberated  inflation'' can  be  useful in  the  context of  physical
theories that do not have slow-roll potentials.

\begin{acknowledgments}
  It's a  pleasure to thank Robert Brandenberger,  Sean Carroll, Chris
  Gordon, Eugene  Lim, Slava Mukhanov  and Govindan Rajesh  for useful
  comments and  discussions.  This work  has been supported by  the US
  DOE grant DE-FG02-90ER40560.
\end{acknowledgments}

\end{document}